\documentclass[11pt]{article}
\usepackage{latexsym}

\def\be{\begin{equation}}
\def\ee{\end{equation}}
\def\bea{\begin{eqnarray}}
\def\eea{\end{eqnarray}}

\def\fn{\footnote}

\def\case#1/#2{\textstyle\frac{#1}{#2}}

\begin{document}
\begin{titlepage}

\vspace{1.4in}

\begin{center}
\Large\bf
{Embeddings in Spacetimes Sourced by Scalar Fields}\normalfont\normalsize
\\
\vspace{1in}\large{E. Anderson$^1$, F. Dahia$^2$, J. E. Lidsey$^3$
and C. Romero$^4$}
\\
\normalsize
\vspace{1.4in}

{$^1$Astronomy Unit, School of Mathematical Sciences, Queen Mary, 
Mile End Road, London E1 4NS, U.K. }

\vspace{.3in}

{$^2$Departamento de F\'{i}sica, Universidade Federal da Para\'{i}ba, C. 
Postal 5008, Jo\~ao Pessoa, PB 58059--970, Brazil}

\vspace{1in}

\begin{abstract}
The extension of the Campbell-Magaard embedding theorem to general 
relativity with 
minimally-coupled scalar fields is formulated and proven.  The result 
is applied 
to the case of a self-interacting scalar field for which new embeddings 
are found, 
and to Brans--Dicke theory.  The relationship between Campbell--Magaard 
theorem 
and the 
general relativity Cauchy and initial value problems is outlined.
\end{abstract}

\end{center}

\end{titlepage}

\section{Introduction}

There is currently a high level 
of interest in higher--dimensional theories of 
gravity, motivated in part by recent developments in both string theory and 
early--universe cosmology. There is a growing body of evidence \cite{WT} 
supporting the conjecture that the five perturbative, 10-d string theories may 
correspond to different limiting cases of a more fundamental, non--perturbative, 
11-d  `M--theory', which reduces to 11-d supergravity in the infra--red limit. 
(For recent reviews, see, e.g., Refs. \cite{stringreview}). 
Inspired by these advances, the proposal that our observable universe may be 
regarded as a domain wall or `brane' that is embedded in a higher-dimensional 
space--time has recently become popular in early--universe cosmology 
\cite{BW,HW,lukas,RS}. 
In the braneworld scenario, the back reaction of the brane results in a  
higher--dimensional geometry that is non--factorizable. Consequently this 
scenario is clearly distinct from the standard Kaluza--Klein compactification 
scheme. Indeed, in the five--dimensional Randall--Sundrum models, the extra 
dimension need not be compact \cite{RS}. 

An important question that arises in such braneworld scenarios is the 
relationship between the geometry of the apparent, lower--dimensional world and 
that of the embedding, higher--dimensional space--time.   
It is therefore important, in view of the above discussion, to develop embedding theorems that 
enable such questions to be concretely addressed (for an overview
and extensive bibliography, see \cite{PT}). An important theorem 
is due to Campbell and Magaard (CM) \cite{campbell, magaard} and  states 
that 
any $n$-dimensional, (semi--)Riemannian manifold $\left(M^{n},g\right)$ can be 
locally and isometrically embedded in an $(n+1)$--dimensional manifold 
$\left( N^{n+1},\tilde{g}\right)$, where the Ricci curvature of $N^{n+1}$ 
vanishes \cite{campbell,magaard}. The theorem was suggested by Campbell \cite{campbell} and a proof 
was later offered by Magaard \cite{magaard}. 
The theorem has been discussed in a number of contexts in the literature 
\cite{romero,maia,othercampbell,lidsey1,ind1,agnese,lidsey,dahia1,dahia2,searha}. 

Recently, we showed \cite{lidsey, dahia1} how to extend this theorem to the 
class of embeddings where $\left( N^{n+1},\tilde{g} \right)$ is an Einstein 
space, with a non--zero Ricci tensor that is directly proportional to the 
metric, $\tilde{g}$. Specific classes of embeddings, such as those of Einstein 
spaces within Einstein spaces, were established \cite{lidsey}. 

Einstein spaces can be viewed as solutions to General Relativity (GR), where 
the Ricci curvature is generated by a particular source -- the cosmological 
constant.  A natural question to address is whether further extensions of the 
CM theorem are possible and, in particular, to investigate embeddings in spaces 
that are sourced by dynamical matter fields. One of the simplest models for 
matter is a scalar field. In this paper, we extend the CM analysis to include a 
minimally-coupled scalar field  with a general potential energy arising through 
its self--interactions, and for Brans--Dicke theory \cite{brans}.  

The paper is organized as follows. In Sec 2 we extend the CM theorem to spaces 
sourced by a scalar field, identifying the mathematical conditions that must be 
satisfied by such a field in order for the embedding of a given manifold to be 
possible in principle.  In Sec 3.1, it is shown that minimally-coupled scalar 
self--interacting scalar fields satisfy these conditions.  Examples of such 
embeddings are given in Sec 3.2.  The Brans--Dicke field is then considered in 
Sec 3.3.  We then outline the clarifying relation between the CM theorem and 
the GR Cauchy \cite{Wald, Bruhat, HawkingEllis} and initial value 
\cite{Lich, Yorkint, YCB} problems in Sec 4.  We conclude in Sec 5.

\section{Campbell--Magaard Theorem with Scalar Field}

The proof of the Campbell-Magaard (CM) theorem\cite{campbell,magaard,romero} 
and of its extension to the case when the embedding manifold is an Einstein 
space \cite{lidsey,dahia1} follows a scheme similar to the methods employed 
when investigating the GR Cauchy Problem, i.e., once the initial conditions 
for the metric in a 3-dimensional hypersurface are given, one would like to 
know whether the Einstein field equations (EFE's) with a non--trivial source 
admit a unique solution. 

In GR, the space--time metric is determined by the Einstein equations
\begin{equation}
\label{GT}
G_{\mu \nu }=-\kappa T_{\mu \nu },
\end{equation}
where $\kappa$ is the Einstein constant and $T_{\mu \nu }$ is the 
energy-momentum tensor, which is a function of the matter fields and the metric.
We consider a scalar field $\bar{\chi}$ defined in a semi-Riemannian
manifold\footnote{In this paper, Latin and Greek indices run from $1$ to $n$ 
and $1$ to $n+1$, respectively.} $(N^{n+1}, \tilde{g}_{\alpha\beta}).$ 
We assume that the energy-momentum tensor of this field is an analytic 
function of the field $\bar{\chi}$, its first derivatives and the metric 
tensor $\tilde{g}_{\alpha \beta }:$ 
\begin{equation}
\tilde{T}_{\mu \nu }=\tilde{T}_{\mu \nu }\left( \bar{\chi},\frac{\partial
\bar{\chi}}{\partial x^{\alpha }},\tilde{g}_{\alpha \beta }\right) ,
\label{T}
\end{equation} 
Let us choose a coordinate system in which the metric 
$\tilde{g}_{\alpha\beta }$ has the line element
\begin{equation}
ds^{2}=\bar{g}_{ik}dx^{i}dx^{k}+\varepsilon \bar{\phi}^{2}dy^{2}
\label{3ds2}
\end{equation}
where $y=x^{n+1},\bar{\phi}=\bar{\phi}\left( x^{1},...,x^{n},y\right)$ 
and $\bar{g}_{ik} = \bar{g}_{ik} \left( x^{1},...,x^{n},y\right)$
and $\varepsilon =\pm 1.$

We suppose that the evolution of $\bar{\chi}$ is governed by a second-order 
p.d.e  which may be written in the form 
\begin{equation}
\frac{\partial ^{2}\bar{\chi}}{\partial y^{2}}=P\left( \bar{\chi},\frac{
\partial \bar{\chi}}{\partial x^{i}},\frac{\partial \bar{\chi}}{\partial y},
\bar{g}_{\alpha \beta },\frac{\partial \bar{g}_{\alpha \beta }}{\partial
x^{i}},\frac{\partial \bar{g}_{\alpha \beta }}{\partial y},\frac{\partial
^{2}\bar{\chi}}{\partial x^{i}\partial x^{k}}\right) ,  
\label{phi}
\end{equation}
where $P$ is analytic with respect to each of its arguments.  We also make 
the physically reasonable assumption of energy-momentum conservation:
\begin{equation} 
\tilde{\nabla}_{\mu }\tilde{T}^{\mu \nu }=0.  
\label{divT}
\end{equation}
$\tilde{\nabla}_{\mu}$ is the covariant derivative with respect to 
$\tilde{g}_{\mu\nu}$. 

We now claim that if conditions (\ref{T}), (\ref{phi}) and (\ref{divT}) 
hold, it is possible to locally embed any $n$-dimensional, (semi--) Riemannian 
manifold $(M^{n}, g) $ in a $(n+1)$-dimensional 
space sourced by the scalar field $\bar{\chi}$. 

In the coordinates (\ref{3ds2}), the EFE's take the form  
\begin{eqnarray}
\tilde{R}_{ik} &=&\bar{R}_{ik}+\varepsilon \bar{g}^{jm}\left( \bar{\Omega}%
_{ik}\bar{\Omega}_{jm}-2\bar{\Omega}_{jk}\bar{\Omega}_{im}\right) -\frac{%
\varepsilon }{\bar{\phi}}\frac{\partial \bar{\Omega}_{ik}}{\partial y}
\nonumber \quad +\frac{1}{\bar{\phi}}\bar{\nabla}_{i}\bar{\nabla}_{k}\bar{\phi}%
=-\kappa \left( \tilde{T}_{ik}-\frac{1}{n-1}g_{ik}\tilde{T}\right) ,
\label{3Rik} \\
\tilde{R}_{i}^{y} &=&\frac{\varepsilon }{\bar{\phi}}\bar{g}^{jk}\left( \bar{%
\nabla}_{j}\bar{\Omega}_{ik}-\bar{\nabla}_{i}\bar{\Omega}_{jk}\right)
=-\kappa \tilde{T}_{i}^{y},  
\label{3Rin+1} \\
\tilde{G}_{y}^{y} &=&-\frac{1}{2}\bar{g}^{ik}\bar{g}^{jm}\left( \bar{R}%
_{ijkm}+\varepsilon \left( \bar{\Omega}_{ik}\bar{\Omega}_{jm}-\bar{\Omega}%
_{jk}\bar{\Omega}_{im}\right) \right) =-\kappa \tilde{T}_{y}^{y},
\label{3Gn+1}
\end{eqnarray}
where   
\begin{equation}
\label{omega}
\bar{\Omega}_{ik}=-\frac{1}{2\bar{\phi}}\frac{\partial \bar{g}_{ik}}{%
\partial y}
\end{equation}
and the barred terms are calculated with the metric $\bar{g}_{ik}
$ induced on the hypersurface $\Sigma _{c}$ of constant $y = c$. 

We now state the following lemma:

{\bf Lemma 1.} {\it Let the functions }$\bar{g}_{ik}\left(
x^{1},...,x^{n},y\right) ${\it , }$\bar{\phi}\left( x^{1},...,x^{n},y\right) 
${\it \ and }$\bar{\chi}\left( x^{1},...,x^{n},y\right) $ {\it be analytic
at }$\left( 0,...,0\right) \in \Sigma _{0}\subset {\Re}^{n+1}$.{\it \
Assume that the following conditions hold }

{\it i)} $\bar{g}_{ik}=\bar{g}_{ki};$

{\it ii)} $\det \left( \bar{g}_{ik}\right) \neq 0;$

{\it iii)} $\bar{\phi}\neq 0${\it .}

\noindent {\it Assume further that }$\bar{g}_{ik}$ {\it and $\bar{\chi}$ 
satisfy the equations (%
\ref{phi}) and (\ref{3Rik}) in the open set }$U\subset {\Re}^{n+1}${\it \
which contains }$0\in {\Re}^{n+1}$ {\it and (\ref{3Rin+1}) and (\ref
{3Gn+1}) at }$\Sigma _{0}.${\it \ Then, }$\bar{g}_{ik}${\it , }$\bar{\phi}$ 
{\it and }$\bar{\chi}$ {\it satisfy (\ref{3Rin+1}) and (\ref{3Gn+1}) in a
neighborhood of }$0\in ${\it \ }${\Re}^{n+1}.$

{\it\bf Proof. }The key point of the proof is given by (\ref{divT}). First,
let us define the tensor $\tilde{F}_{\alpha \beta }=\tilde{G}_{\alpha \beta
}+\kappa \tilde{T}_{\alpha \beta }.$ By assumption, $\bar{\chi}$ satisfies (%
\ref{phi}) in a neighborhood $V\subset $ ${\Re}^{n+1}$ of $0\in {\Re}%
^{n+1},$ whence (\ref{divT}) also holds in $V.$ It 
then follows that $\tilde{F}%
_{\alpha \beta }$ has vanishing divergence, so
\[
\frac{\partial \tilde{F}_{\beta }^{y}}{\partial y}=-\frac{\partial \tilde{F}%
_{\beta }^{i}}{\partial x^{i}}-\tilde{\Gamma}_{\mu \lambda }^{\mu }\tilde{F}%
_{\beta }^{\lambda }+\tilde{\Gamma}_{\lambda \beta }^{\mu }\tilde{F}_{\mu
}^{\lambda }.
\]
On the other hand, by expressing the Einstein tensor in terms of the Ricci
tensor, we can write
\[
\tilde{F}_{k}^{i}=\tilde{R}_{k}^{i}-\delta _{k}^{i}\left( \tilde{R}_{j}^{j}+%
\tilde{G}_{y}^{y}\right) +\kappa \tilde{T}_{k}^{i}  .
\]
Again, by assumption, the equation 
\begin{eqnarray}
\tilde{R}_{k}^{i}=-\kappa \left[ \tilde{T%
}_{k}^{i}-\frac{\delta _{j}^{i}}{n-1}\left( \tilde{T}_{j}^{j}+\tilde{T}%
_{y}^{y}\right) \right] 
\nonumber
\end{eqnarray}
holds in $V\subset $ ${\Re}^{n+1}$ and it then follows that 
$\tilde{F}_{k}^{i}=-\delta _{k}^{i}\tilde{F}_{y}^{y}$
in $V$.
After some
algebra we may then deduce that 
\begin{eqnarray*}
\frac{\partial \tilde{F}_{y}^{y}}{\partial y} &=&-\varepsilon \bar{\phi}^{2}%
\bar{g}^{ij}\frac{\partial \tilde{F}_{i}^{y}}{\partial x^{j}}-2\tilde{\Gamma}%
_{iy}^{i}\tilde{F}_{y}^{y}+\left( -\varepsilon \frac{\partial \left( \bar{%
\phi}^{2}\bar{g}^{ij}\right) }{\partial y^{j}}-\varepsilon \bar{\phi}^{2}%
\bar{g}^{ij}\tilde{\Gamma}_{kj}^{k}+\tilde{\Gamma}_{yy}^{i}\right) \tilde{F}%
_{i}^{y} \\
\frac{\partial \tilde{F}_{i}^{y}}{\partial y} &=&\frac{\partial \tilde{F}%
_{y}^{y}}{\partial x^{i}}+2\tilde{\Gamma}_{yi}^{y}\tilde{F}_{y}^{y}+\left(
\tilde{\Gamma}_{yi}^{k}+\varepsilon \bar{\phi}^{2}\bar{g}^{kj}\tilde{\Gamma}%
_{ij}^{y}-\tilde{\Gamma}_{y\mu }^{\mu }\delta _{i}^{k}\right) \tilde{F}%
_{k}^{y} 
\end{eqnarray*}
and since (\ref{3Rin+1}) and (\ref{3Gn+1}) hold
at the hypersurface $\Sigma _{0}$, 
it follows that $\tilde{F}_{\beta }^{y}=0$ and hence 
that $\left. 
\partial \tilde{F}_{\beta }^{y}/\partial y \right| _{y=0}=0.$ 

It is not difficult to show by 
mathematical induction that all the derivatives
of $\tilde{F}_{\beta }^{y}$ (to any order)  vanish at $y=0.$ As $\tilde{F}%
_{\beta }^{y}$ is analytic,  
we may therefore conclude that $\tilde{F}_{\beta }^{y}=0$ in an
open set of ${\Re}^{n+1}.$ Thus, (\ref{3Rin+1}) and (\ref{3Gn+1}) also
hold in an open set of ${\Re}^{n+1}$ which includes the origin
and this
proves the lemma. $\Box$

To proceed, we must now establish that the 
solutions 
$\bar{g}_{ik}$, $\bar{\phi}$ and $\bar{\chi}$ 
do indeed exist. With this in mind, 
let us now recall the Cauchy-Kowalewski (CK) 
theorem \cite{courant}:

{\bf Theorem (Cauchy-Kowalewski). }{\it Consider the set of partial
differential equations}$:$%
\begin{equation}
\frac{\partial ^{2}u^{A}}{\partial \left( y^{n+1}\right) ^{2}}=F^{A}\left(
y^{\alpha },u^{B},\frac{\partial u^{B}}{\partial y^{\alpha }},\frac{\partial
^{2}u^{B}}{\partial y^{\alpha }\partial y^{i}},\right) ,\qquad A=1,...,m ,
\label{CK}
\end{equation}
{\it where }$u^{1},..,u^{m}$ {\it are }$m$ {\it unknown functions of the }$%
n+1$ {\it variables }$y^{1},...,y^{n},y^{n+1},$ $\alpha =1,...,n+1,$ $%
i=1,..,n,$ $B=1,...,m.$ {\it \ Also, let }$\xi ^{1},...,\xi ^{m},\eta
^{1},...,\eta ^{m},$ {\it be functions of the variables }$y^{1},...,y^{n},$
{\it and be analytic at }$0\in ${\it \ }${\Re}^{n}.${\it \ If the functions }$%
F^{A}$ {\it are analytic with respect to each of their arguments around the
values evaluated at the point }$y^{1}=...=y^{n}=0,${\it \  there exists
a unique solution of equations (\ref{CK}) which is analytic at }$0\in {\Re%
}^{n+1}$ {\it and that satisfies the initial conditions} 
\begin{eqnarray}
u^{A}\left( y^{1},...,y^{n},0\right)  &=&\xi ^{A}\left(
y^{1},...,y^{n}\right)  \\
\frac{\partial u^{A}}{\partial y^{n+1}}\left( y^{1},...,y^{n},0\right) 
&=&\eta ^{A}\left( y^{1},...,y^{n}\right) ,\qquad A=1,...,m.
\end{eqnarray}

After solving (\ref{3Rik}) 
for the second-order derivative of $\bar{g}_{ik}$
with respect to $y$ we find that  
\begin{eqnarray}
\frac{\partial ^{2}\bar{g}_{ik}}{\partial y^{2}} &=&-2\varepsilon \kappa 
\bar{\phi}^{2}\left( T_{ik}-\frac{1}{n-1}g_{ik}T\right) +\frac{1}{\bar{\phi}}%
\frac{\partial \bar{\phi}}{\partial y}\frac{\partial \bar{g}_{ik}}{\partial y%
}-\frac{1}{2}\bar{g}^{jm}\left( \frac{\partial \bar{g}_{ik}}{\partial y}%
\frac{\partial \bar{g}_{jm}}{\partial y}-2\frac{\partial \bar{g}_{im}}{%
\partial y}\frac{\partial \bar{g}_{jk}}{\partial y}\right)   \nonumber \\
&&-2\varepsilon \bar{\phi}\left( \frac{\partial ^{2}\bar{\phi}}{\partial
x^{i}\partial x^{k}}-\frac{\partial \bar{\phi}}{\partial x^{j}}\bar{\Gamma}%
_{ik}^{j}\right) -2\varepsilon \bar{\phi}^{2}\bar{R}_{ik}.  \label{3eqgik}
\end{eqnarray}
Due to the symmetry of the tensors, $\bar{g}_{ik}$ and $T_{ik}$, we can
rewrite (\ref{3eqgik}) in terms of the components of $g_{ik}$ with $%
i\leq k.$ This equation, together with the field equation (\ref{phi}), form a
set of $1+ n(n+1)/2$ p.d.e's for the $1+ n\left( n+1\right) /2$
unknown functions,   $\bar{g}_{ik}$ $\left( i\leq k\right) $ and $\bar{%
\chi}.$ (Note that $\bar{\phi}$ is a nonzero analytic function 
that is treated as a known).

Thus, the p.d.e system we have just obtained has the
canonical form of (\ref{CK}) and, moreover, it satisfies all 
of the conditions required for the use of the CK theorem. Indeed, by virtue of
the  properties (\ref{T}) and (%
\ref{phi}) imposed on $T_{\alpha \beta }$ and $P$, the right-hand side of the
equations is comprised of functions of the variables  
$$
x^{1},..,x^{n},y;\bar{g}_{ik},\bar{\chi},\frac{\partial \bar{g}_{ik}}{%
\partial x^{j}},\frac{\partial \bar{g}_{ik}}{\partial y},\frac{\partial \bar{%
\chi}}{\partial x^{j}},\frac{\partial \bar{\chi}}{\partial y};\frac{\partial
^{2}\bar{g}_{ik}}{\partial x^{j}\partial x^{m}},\frac{\partial ^{2}\bar{\chi}%
}{\partial x^{j}\partial x^{m}},
$$
which are analytic with respect to each of their arguments at 
$$
x^{1}=0,..,x^{n}=0,y=0;\left. \bar{g}_{ik}\right| _{0},\left. \bar{\chi}%
\right| _{0},\left. \frac{\partial \bar{g}_{ik}}{\partial x^{j}}\right|
_{0},\left. \frac{\partial \bar{g}_{ik}}{\partial y}\right| _{0},\left. 
\frac{\partial \bar{\chi}}{\partial x^{j}}\right| _{0},\left. \frac{\partial
\bar{\chi}}{\partial y}\right| _{0};\left. \frac{\partial ^{2}\bar{g}_{ik}}{%
\partial x^{j}\partial x^{m}}\right| _{0},\left. \frac{\partial ^{2}\bar{\chi%
}}{\partial x^{j}\partial x^{m}}\right| _{0},
$$ 
if $\left| \bar{g}_{ik}\right| _{0}\neq 0.$ Therefore, given analytic
initial conditions
\be
\begin{array}{c}
\bar{g}_{ik}\left( x^{1},..,x^{n},0\right)  = g_{ik}\left(
x^{1},..,x^{n}\right) 
\mbox{ } , \mbox{ } \mbox{ } 
\frac{\partial \bar{g}_{ik}}{\partial y}\left( x^{1},..,x^{n},0\right)  = 
\bar{\phi}\left( x^{1},..,x^{n},0\right) \Omega _{ik}
\left(
x^{1},...,x^{n}
\right)   
\label{3cih} \\
\bar{\chi}\left( x^{1},..,x^{n},0\right)  =\xi 
\left(
x^{1},..,x^{n}
\right) 
\mbox{ } , \mbox{ } \mbox{ } 
\frac{\partial \bar{\chi}}{\partial y}\left( x^{1},..,x^{n},0\right) = 
\eta \left( x^{1},..,x^{n}\right) ,  \label{3cidphi}
\end{array}
\ee
which satisfy $\left| g_{ik}\right| \neq 0,$ there exists a unique set of
functions $\bar{g}_{ik}$ and $\bar{\chi}$ which solve the equations (\ref
{phi}) and (\ref{3Rik}) which are analytic at the origin. It should be noted
that an important feature of these solutions is the property 
$\left| \bar{g}_{ik}\right| \neq 0$ in some neighborhood of $0\in {\Re}^{n+1}.$

If we take the initial conditions $g_{ik}$ as being the metric components of
a (semi--) Riemannian space $M^{n}$ written in some coordinate system, then we
can state the following theorem:

{\bf Theorem 1 }{\it Let }$M^{n}$ {\it be a n-dimensional, semi--Riemaninan
manifold with metric given by} $ds^{2}=g_{ik}dx^{i}dx^{k}$,
{\it \ in a coordinate system }$\left\{ x^{i}\right\} $ {\it of }$M^{n}.$ 
{\it Let }$p\in M^{n}$ {\it have coordinates }$x_{p}^{1}=...=x_{p}^{n}=0.$
{\it Then }$M^{n}$ {\it has a local isometric and analytic embedding (at
the point }$p$) {\it in an (n+1)-dimensional space 
sourced by any arbitrary scalar
field }$\bar{\chi}$ {\it \ that is 
characterized by the properties (\ref{T}), (\ref
{phi}) and (\ref{divT})} {\it iff there exist functions }$\Omega
_{ik}\left( x^{1},...,x^{n}\right) ${\it \ }$\left( i,k=1,..,n\right) ,$ $%
\xi \left( x^{1},..,x^{n}\right) $, $\eta \left( x^{1},..,x^{n}\right) $
{\it and} $\phi \left( x^{1},..,x^{n}\right) \neq 0$ {\it that are analytic
at }$0\in {\Re}^{n}$ {\it such that }
\begin{eqnarray}
\Omega _{ik} &=&\Omega _{ki}  \label{3eq1} \\
g^{jk}\left( \nabla _{j}\Omega _{ik}-\nabla _{i}\Omega _{jk}\right) 
&=&-\varepsilon \kappa \phi T_{i}^{y}\left( \xi ,\eta ,g_{ij}\right)
\label{3eq2} \\
g^{ik}g^{jm}\left( R_{ijkm}+\varepsilon \left( \Omega _{ik}\Omega
_{jm}-\Omega _{jk}\Omega _{im}\right) \right)  &=&2\kappa T_{y}^{y}\left(
\xi ,\eta ,g_{ij}\right) .  \label{3eq3}
\end{eqnarray}

{\it\bf Proof.} ($\Rightarrow$) If $M^{n}$
has an embedding in some arbitrary space sourced by a scalar field, then it
can be proved \cite{dahia1,tese} 
that there exists a coordinate system in which the metric of
the embedding space has the form 
$ds^{2}=\bar{g}_{ik}dx^{i}dx^{k}+\varepsilon \bar{\phi}^{2}dy^{2}$, 
where the analytic functions $\bar{g}_{ik}\left( x^{1},...,x^{n},y\right) $
and $\bar{\phi}\left( x^{1},...,x^{n},y\right) $ 
are such that $\bar{\phi}\left(
x^{1},...,x^{n},y\right) \neq 0$ and 
$\bar{g}_{ik}\left(
x^{1},...,x^{n},0\right) =g_{ik}\left( x^{1},...,x^{n}\right) $ in an open
set of ${\Re}^{n}$ which contains the origin. Given that the embedding
space is, by assumption, generated by a 
scalar field $\bar{\chi},$ it follows that $\bar{g%
}_{ik}$ and $\bar{\phi}$ necessarily satisfy the equations (\ref{phi}), (\ref
{3Rik}), (\ref{3Rin+1}) and (\ref{3Gn+1}) for some field $\bar{\chi}\left(
x^{1},...,x^{n},y\right) $ in a neighborhood of $0\in {\Re}^{n+1}.$ In
particular, the equations (\ref{3Rin+1}) and (\ref{3Gn+1}) hold for $y=0.$
Therefore, the functions $\Omega _{ik}\left( x^{1},...,x^{n}\right) ${\it \ }%
$\left( i,k=1,..,n\right) ,$ $\xi \left( x^{1},..,x^{n}\right) $, $\eta
\left( x^{1},..,x^{n}\right) $ and $\phi \left( x^{1},..,x^{n}\right) $,
as 
defined by 
\begin{eqnarray}
\Omega _{ik} &=&\bar{\Omega}_{ik}\left( x^{1},..,x^{n},0\right) ,\quad \xi =%
\bar{\chi}\left( x^{1},...,x^{n},0\right)
\eta  = \left. \frac{\partial \bar{\chi}}{\partial y}\right| _{y=0},\quad
\quad \phi \left( x^{1},..,x^{n}\right) =\bar{\phi}\left(
x^{1},...,x^{n},0\right) 
\end{eqnarray}
satisfy the equations (\ref{3eq1}), (\ref{3eq2}) and (\ref{3eq3}).

($\Leftarrow$) Suppose that there exist functions $\Omega
_{ik}\left( x^{1},...,x^{n}\right) ,$ $\xi \left( x^{1},..,x^{n}\right) $, $%
\eta \left( x^{1},..,x^{n}\right) $ and $\phi \left( x^{1},..,x^{n}\right)
\neq 0$ which satisfy (\ref{3eq1}), (\ref{3eq2}) and (\ref{3eq3}). Choose an
analytic function $\bar{\phi}\left( x^{1},...,x^{n},y\right) \neq 0$ such
that $\bar{\phi}\left( x^{1},...,x^{n},0\right) =\phi \left(
x^{1},..,x^{n}\right) .$ By virtue of the CK theorem, there
exists a unique set of analytic functions $\bar{g}_{ik}\left(
x^{1},...,x^{n},y\right) $ and $\bar{\chi}\left( x^{1},...,x^{n},y\right) $
that satisfy  the equations (\ref{phi}) and (\ref{3Rik}) and the initial
conditions (\ref{3cih}).
Since, by assumption, the initial conditions satisfy  equations (\ref{3eq1}%
), (\ref{3eq2}) and (\ref{3eq3}), equations  
(\ref{3Rin+1}) and (\ref{3Gn+1}) are satisfied at $y=0$
by $\bar{g}_{ik},\bar{\phi}$ and $\bar{%
\chi}$. It follows from
Lemma 1 that $\bar{g}_{ik},\bar{\phi}$ and $\bar{\chi}$ also 
satisfy equations (\ref{phi}),
(\ref{3Rik}) and (\ref{3Gn+1}) in an open set of ${\Re}^{n+1}
$ which contains the origin. We conclude, therefore,
that the $(n+1)$-dimensional manifold
whose line element (\ref{3ds2}) is expressed in terms
of the solutions $\bar{g}_{ik}$
and $\bar{\phi}$ is a space generated by a scalar field. 
Thus, the
(semi--)Riemannian manifold $\left( M^{n},g\right) $ can 
indeed be embedded in a space sourced by a scalar field. $\Box $

According to Theorem 1, the existence of solutions to equations (\ref
{3eq1}), (\ref{3eq2}) and (\ref{3eq3}) is sufficient to 
ensure that the local,
analytic embedding of $M^{n}$ is possible.
The proof that these solutions do in fact exist
consists in showing that equations 
(\ref{3eq1}), (\ref{3eq2}) and (\ref{3eq3}) can be
expressed in the canonical form required by the CK 
theorem (here in its
first-derivative version). This can be done by following 
the method presented in \cite{romero} with no significant modifications, 
and so we omit the details here. The idea is to use (\ref{3eq3}) to 
isolate $\Omega_{11}$ which is then substituted into (\ref{3eq2}) which 
is to be regarded as a system of p.d.e's for $n$ unknown functions: the 
$n - 1$ $\Omega_{1k}$ for $k\geq 2$ and one other component of $\Omega$, 
which we call $\Theta$.
These p.d.e's are in the correct canonical form to employ the CK theorem.
Thus, if the analytic functions $g_{ik}$ are given, there exist
analytic functions $\Omega _{ik}$ which solve 
equations (\ref{3eq1}), (\ref{3eq2}) and
(\ref{3eq3}) and this leads us to the following theorem:

{\bf Theorem 2.} {\bf \ }{\it Let }$M^{n}$ $(n>1)$ {\it be a piece of a 
(semi--)Riemannian space with line element}
$ds^{2}=g_{ik}dx^{i}dx^{k}$,
{\it expressed in a coordinate system which covers a neighborhood of a point
}$p\in M^{n}$ {\it whose coordinates are }$x_{p}^{1}=...=x_{p}^{n}=0.$ {\it 
If }$g_{ik}$ {\it are analytic functions at }$0\in {\Re}^{n},${\it \ then
}$M^{n}${\it \ can be embedded at }$p$ {\it in some 
(n+1)--dimensional space sourced by any arbitrary
scalar field satisfying the conditions}
{\it (\ref{T}), (\ref{phi}) and (\ref{divT}). Moreover, the line element of
the embedding space is unique if the arbitrary functions to be chosen obey the
following conditions:}

{\it i) the }$[n ( n-1)/2]-1${\it \ functions }$\Omega _{ik}
${\it \ }$\left( i\leq k,i>1, \mbox{ and excluding the 
component } \Theta \right) ${\it \ are analytic at }$0\in \Re^{n};$

{\it ii) the }$n${\it \ functions }$\Omega _{1k}\left(
0,x^{2},...,x^{n}\right) =f_{k}\left( x^{2},...,x^{n}\right) ${\it \ }$%
\left( k>1\right) ${\it \ and }$\Theta \left(
0,x^{2},...,x^{n}\right) =f_{1}\left( x^{2},...,x^{n}\right) ${\it \ are
analytic at }$0\in \Re^{n-1},${\it \ with the coefficient of }$\Omega_{11}$ 
{\it in (\ref{3eq3}) nonzero (to permit the elimination of } $\Omega_{11}$ 
{\it in order to set up the p.d.e system)}\normalfont,

{\it iii) a function }$\phi \left( x^{1},...,x^{n+1}\right) \neq 0${\it %
, analytic at }$0\in {\Re}^{n+1},${\it \ is chosen;}

{\it iv) two functions }$\xi \left( x^{1},...,x^{n}\right) $ {\it and} $\eta
\left( x^{1},...,x^{n}\right) ${\it , analytic at }$0\in {\Re}^{n},$ {\it %
are chosen.}

Note that when we refer  to a space generated by a scalar
field we have in mind a non--trivial solution of the Einstein--scalar system. 
In other words, we are implicitly considering a solution such that the
associated energy--momentum tensor is nonzero. The exclusion of trivial
solutions is possible because the initial conditions of the field (the
functions $\xi $ and $\eta $) are arbitrary 
and this implies that they can be chosen in such a way
that $T_{\alpha \beta }\neq 0$ 
at $\Sigma
_{0}$
(for some $\alpha $ and $\beta $). 
Consequently, continuity requirements imply that the energy--momentum tensor
does not vanish in some neighborhood of $0\in {\Re}^{n+1}.$

We proceed in the following Section to employ this theorem to 
establish specific classes of embeddings. 

\section{Applications}

\subsection{Self--interacting scalar fields}

In this section we consider a real scalar field, $\chi$, 
minimally coupled to Einstein gravity and self--interacting through a 
potential, $W (\chi )$. We assume that the potential is a well--behaved, 
analytic function of the field. 
The energy--momentum tensor of such a field is given by 
\begin{equation}
\label{enmom}
T_{\alpha \beta }=\nabla _{\alpha }\chi \nabla _{\beta }\chi -\frac{1}{2}
g_{\alpha \beta }\left( \nabla ^{\gamma }\chi \nabla _{\gamma }\chi
\right) - g_{\alpha\beta} W (\chi ). 
\end{equation}
It easily follows that the energy--momentum tensor (\ref{enmom}) has 
vanishing divergence if $\chi$ solves the field equation
\begin{equation}
\label{KG}
\nabla ^{\alpha }\nabla _{\alpha }\chi -\frac{dW}{d\chi} =0.  
\end{equation}
In the coordinate system (\ref{3ds2}) the field equation (\ref{KG}) takes
the form of equation (\ref{phi}) with
\begin{equation}
P=\varepsilon \phi ^{2}\left( -g^{ik}\frac{   \partial ^{2}\chi   }
{   \partial x^{i}  \partial x^{k}   } - g^{\alpha \beta }
\Gamma _{\alpha \beta }^{\gamma }\frac{ \partial \chi }
{  \partial x^{\gamma }  }+\frac{  dW  }{  d\chi  } \right) .
\end{equation}
Note that the function 
P which depends on $\chi$ ,$\frac{  \partial \chi }
{ \partial x^{i} }$, $\frac{ \partial \chi }{\partial y}$, $g_{\alpha \beta }$, 
$\frac{  \partial g_{\alpha \beta }  }{  \partial x^{i}  }$, 
$\frac{  \partial g_{\alpha \beta }  }{  \partial y  }$, and 
$\frac{  \partial ^{2}\chi }{  \partial x^{i}\partial x^{k}  }$, 
is analytic with respect to each of these arguments. 

Since conditions (\ref{T}), (\ref{phi}) and (\ref
{divT}) are satisfied by a minimally coupled, 
self--interacting scalar field, we may conclude the following:

{\bf Corollary 1.} {\bf \ }{\it Let }$M^{n}(n>1)$ {\it be a piece of a 
(semi--)Riemannian space with line element} 
$ ds^{2}=g_{ik}dx^{i}dx^{k}$,
{\it expressed in a coordinate system which covers a neighborhood of a point 
}$p\in M^{n}$ {\it whose coordinates are }
$x_{p}^{1}=...=x_{p}^{n}=0.$ {\it %
If }$g_{ik}$ {\it are analytic functions at }$0\in {\Re}^{n},${\it \ then 
}$M^{n}${\it \ can be embedded at }$p$ {\it in some 
(n+1)--dimensional space generated by any arbitrary, minimally 
coupled, self--interacting scalar field.}

Note that $M^{n}$ is truly Riemannian for $\varepsilon = -1$ and 
semi--Riemannian for $\varepsilon =+ 1$. 

\subsection{Examples of Embeddings with Self-Interacting Scalars}

We now employ this result to construct embeddings for a class of space--times 
into higher-dimensional space--times sourced by such a scalar field. 
We begin with the {\em ansatz} 
\begin{equation}
\label{ansatz}
\bar{\Omega}_{ij} = C\bar{g}_{ij} ,
\end{equation}
where $C=C( x^{\alpha})$ is a scalar function 
of the coordinates of the embedding metric. We 
further assume that the scalar 
field is independent of the embedded metric coordinates, 
i.e., $\chi =\chi (y)$, and specify 
$\kappa =1$ and $\varepsilon =1$ for simplicity. 

Substituting (\ref{ansatz}) into (\ref{3eq2}) implies 
that $\nabla_i C =0$ and, consequently, that $C$ must be a function of the extra coordinate $y$ alone. 
Choosing the normal coordinate form for $\phi$: 
\begin{equation}
\label{lapseform}
\phi =1,
\end{equation}
then implies that (\ref{omega}) can be integrated 
to yield the solution
\begin{equation}
\bar{g}_{ij} = a^2 (y) g_{ij}  ,
\end{equation}
where $C \equiv - d \ln a /dy$.

When (\ref{ansatz}) and (\ref{lapseform}) are 
valid, the scalar field equation 
(\ref{KG}) simplifies to 
\begin{equation}
\label{chiform}
\frac{d^2 \chi}{dy^2} +\frac{n}{a} 
\frac{d a}{dy}  \frac{d \chi}{dy} -\frac{dV}{d \chi} 
=0
\end{equation}
and by 
substituting the appropriate components of 
the energy--momentum tensor (\ref{enmom}) into 
(\ref{3Rik}) and (\ref{3Gn+1}), 
we find that 
\begin{eqnarray}
\label{pot1}
\bar{R}_{ik} +\bar{\Omega} \bar{\Omega}_{ik} 
-2 \bar{\Omega}_{im} \bar{\Omega}^m_k 
-\frac{\partial \bar{\Omega}_{ik}}{\partial y} 
= - \frac{2V}{n-1} \bar{g}_{ik},
\\
\label{pot2}
\bar{R} + \bar{\Omega}^2 -\bar{\Omega}^i_j 
\bar{\Omega}^j_i = \left( \frac{d\chi}{d y} \right)^2 -2V.
\end{eqnarray}
Furthermore, subtracting (\ref{pot2}) from the trace of (\ref{pot1}) and 
substituting 
for the {\em ansatz} (\ref{ansatz}) 
implies that 
\begin{equation}
\label{tracesubtract}
\frac{n}{a} \frac{d^2 a}{dy^2} = - 
\left( \frac{d \chi}{d y} \right)^2 - \frac{2V}{n-1}. 
\end{equation}

We may solve (\ref{tracesubtract}) and 
(\ref{chiform}) for an unknown self--interaction potential, 
by  specifying the functional forms of  $a (y)$  and $\chi (y)$ : 
\begin{eqnarray}
\label{afunction}
a \equiv (1 + \lambda y )^p, \\
\label{chifunction}
\chi \equiv q \ln (1+\lambda y ),
\end{eqnarray}
where  $p$, $q$ and $\lambda$ are constants. 
Substituting (\ref{afunction}) and (\ref{chifunction}) 
into (\ref{tracesubtract}) implies that this is consistent if 
the scalar field potential has an exponential (Liouville) form: 
\begin{equation}
\label{liouville}
V = -\lambda^2  \left( \frac{n-1}{2} \right) 
\left[ np (p-1) +q^2 \right] \exp \left( -\frac{2}{q} \chi 
\right), 
\end{equation}
and it then follows that the scalar field 
equation  (\ref{chiform}) is solved 
if 
\begin{equation}
\label{2cases}
(p-1) \left[ q^2 - p(n-1) \right] =0.
\end{equation}

Thus, the embedding is established 
by solving (\ref{pot1}) 
for the two cases, 
where $q^2 =p(n - 1)$ or $p = 1$. 
In the first case, substituting
for (\ref{afunction}), (\ref{chifunction}) and (\ref{liouville}) implies 
that the embedding 
metric, $g_{ij}$, must have vanishing Ricci tensor. 
We conclude, therefore, that there is an embedding of $n$--d  
Ricci--flat manifolds with metric $g_{ij}$, in an $(n+1)$--d  
manifold sourced by a scalar field, with metric given by 
\begin{equation}
\label{expman}
ds^2 = (1 + \lambda y )^p g_{ik} dx^idx^k +dy^2,
\end{equation}
where the scalar field (\ref{chifunction}) 
self--interacts through the exponential potential (\ref{liouville}).  This 
generalizes the 4-d result of 
\cite{Feinstein} to arbitrary dimension. 

In the second case, where $p=1$, the above procedure 
implies that (\ref{pot1}) is solved if 
\begin{equation}
R_{ik} = \frac{(q^2 - n + 1)\lambda^2}{(1 + \lambda y)^2} g_{ik}
\end{equation}
and it follows that the embedded metric 
is an Einstein space with a non--zero Ricci curvature. 
The sign and magnitude of the effective cosmological constant
of the embedded manifold
determine the self--interaction coupling, $q$, of the 
scalar field. Indeed, the potential (\ref{liouville}) 
is negative--definite for this embedding.  
An embedding of this type has been considered within the context of dilatonic braneworld scenarios \cite{Alonso}.  
It is interesting that the functional 
form of the potential (\ref{liouville}) is the same for 
the two different classes of embedding. Moreover, potentials of this form arise in compactified supergravity theories \cite{Lu}.

The above embeddings are specific in the sense that for the assumed form of $\Omega_{ik}$ given in (\ref{ansatz}), 
the embedding is only consistent if the scalar field has an exponential potential.  However, Corollary 1 states 
that an embedding is possible for any analytic potential.  This would require some other form for $\Omega_{ik}$ to be chosen. 

\subsection{Brans--Dicke Theory}

The Brans--Dicke theory of gravity 
\cite{brans} represents a natural extension of GR,
where a nonminimally--coupled (`dilatonic') scalar field 
parametrizes the space--time dependence of Newton's `constant'. 
Nevertheless, it is well-known that this theory is 
conformally equivalent to GR with a minimally--coupled scalar field.   
Therefore, the following corollary to the above theorems is deduced if 
one is prepared to work 
within the context of the conformally-transformed fields. 

{\bf Corollary 2.} {\bf \ \ }{\it Let }$M^{n}(n>1)$ {\it be a piece of a 
(semi--)Riemannian space with line element} $ds^{2}=g_{ik}dx^{i}dx^{k}$,
{\it expressed in a coordinate system which covers a neighborhood of a point
}$p\in M^{n}$ {\it whose coordinates are }$x_{p}^{1}=...=x_{p}^{n}=0.$ {\it %
If }$g_{ik}$ {\it are analytic functions at }$0\in {\Re}^{n},${\it \ then
}$M^{n}${\it \ can be embedded at }$p$ {\it in some }$\left( n+1\right) $%
{\it -dimensional space which is a solution of the vacuum, Brans--Dicke field
equations.\normalfont}

\section{Campbell-Magaard and the GR Cauchy and Initial  Value Problems}

In Sec 2, we referred to 
the close relationship 
between the CM theorem and the GR Cauchy and initial 
value problems (CP and IVP, respectively).   
There is a wealth of literature on the latter problems
within the context of $(3,0) \longrightarrow (3,1)$  
embeddings\footnote{We denote by $(p,q)$ a manifold with 
$p$ spacelike and $q$ timelike coordinates.}, which 
makes it a valuable source of ideas for embedding theorems.  
The first idea is to show how the CM theorem follows from a 
collection of known results.  The second idea will be to question 
whether any of the dimension- or signature-dependent results of the 
GR CP have useful generalizations. 
This will lead to possible limitations of the usual methods of 
constructing higher--dimensional solutions from lower--dimensional ones. 
We begin by outlining the GR CP and IVP.

By use of the $(3 + 1)$ split of the space--time line element\fn{Here, 
$N$ is the lapse, $N^i$ is the shift and $h_{ij}$ is the induced 3-metric.}
\begin{equation}
ds^2 = h_{ij}(Ndx^i + N^idt)(Ndx^j + N^jdt) - (Ndt)^2,
\label{grcpsplit}
\end{equation}
the 10 EFE's may be rewritten as 6 evolution 
equations and the 4 Gauss--Codazzi constraints.  
The mathematics of this split is the same, up to the signature, 
as that of splitting the EFE's into equation 
(6) and  into equations (7) and (8), if one identifies the 
lapse to be $\phi$ and imposes the partial gauge condition 
that the shift be zero.      
The GR CP is then the study of the evolution equations given 
some initial data that obeys the constraints.   

It is standard knowledge \cite{Wald, Bruhat} that these evolution 
equations can be written in the correct form required by
the CK theorem, and so we are guaranteed  
that a solution exists locally and that it is unique.  
Whereas self--consistency requires the evolution equations to
propagate the constraints off the
initial hypersurface, it is immediately evident that this
follows from the Bianchi identities.   

There are additional results about the GR CP having a number of 
physically-desirable features.    
Firstly, it is Hadamard well-posed \cite{courant, Wald} 
so that in addition to the existence of a unique solution, 
the solution depends continuously on the prescribed data.   
Without this, an arbitrarily small change in the
data set could give rise to an arbitrarily large change in the 
form of the solution which prevents physical predictability.   
Secondly, it possesses the `domain of dependence' property, i.e., the 
data can only affect the evolution in regions that are in 
causal contact with that data, 
which is a necessary criterion for the good behaviour of hyperbolic systems
\cite{Wald, HawkingEllis}.
The continuous dependence and domain of dependence properties follow in 
harmonic 
coordinates from Leray's theorem \cite{leray, Wald}, 
which is specific to hyperbolic operators.
Finally, it is standard knowledge that the GR CP is well-posed 
in the presence
of scalar fields, electromagnetism, perfect fluids, Brans--Dicke 
theory and certain higher-derivative theories 
\cite{Wald}.  

The IVP or data construction problem (solving the Gauss--Codazzi 
constraints) is considered to be the most
difficult step in the $(3+1)$ formalism of GR \cite{MTW}. This problem 
is under-determined, 
because there are 12 
unknown functions (the components of the 3-metric $h_{ik}$
and the second fundamental form 
$\Omega_{ik}$), but only 4 equations.  Hawking and Ellis 
\cite{HawkingEllisp232} apparently refer to  
a general result stating that 
if eight of the unknown functions are specified
for a space--time with an arbitrary matter content,  
the  constraint equations may be solved for the 
remaining four.  
However, the work they refer to, Ref. \cite{Bruhat}, 
considers only two methods (the thin sandwich method and the usual 
conformal method), neither of which resemble Magaard's method.  

We now consider the CM theorem piece by piece.  The structure of none of 
the above results are dependent on the dimension.  
Neither the CK theorem nor the method of expressing the GR evolution 
equations in the correct form to invoke the CK theorem 
are affected by the signature.
Furthermore the Bianchi identity which guarantees constraint propagation 
is geometrical and hence equally valid regardless of the signature.  
Hence, the first part of the proof of the CM 
theorem (existence, uniqueness and constraint propagation) follows directly 
from these results holding for the $(3+1)$--d GR CP.   

The second part of the proof, due to Magaard, 
is a subcase of the result in Hawking and 
Ellis \cite{HawkingEllisp232} generalized to arbitrary dimension.     
However, we emphasize that Magaard's 
approach differs from that of the usual conformal data construction 
\cite{Lich, Yorkint, YCB}.  
The former considers the lower-dimensional metric to be a known
function,
but the latter considers it to be  known only up to a conformal factor.
The former is of interest following the new motivation for studying
higher-dimensional space--times, whereas the latter has only been employed 
for three spacelike dimensions using largely signature-dependent 
(that is, elliptical) methods.

If we are to interpret the CM theorem from a physical 
point of view, we are well-motivated to demand that the procedure 
for constructing embeddings is a well-posed problem. 
The types of embeddings of interest are 
$(n,1) \longrightarrow
(n+1,1)$ and $(n,1) \longrightarrow (n,2)$, 
which are forms which are not known to exhibit continuous dependence,
respectively\footnote{There is no simple way of investigating 
whether theorems that
hold for one particular signature hold for any other. This is 
due to significant mathematical differences between elliptic 
and `hyperbolic' operators.  
>From the point of view of the function spaces that underly the analysis,  
Holder spaces are natural for the study of elliptic operators 
while Sobolev spaces are 
natural for the study of hyperbolic operators.}.
The former is termed the `sideways problem for a hyperbolic 
system', and arises in a number of non--relativistic
contexts, as summarized by 
Ames and Straughan \cite{amesstraughan}.  The only general
result known for simple examples of such problems is that, given certain 
bounds, the solution
is Holder-stable in some region \cite{amesisakov}, but  
this is not considered to amount to a sufficiently strong theorem 
to guarantee
continuous dependence. The latter is an ultrahyperbolic problem.
These remain largely unexplored since they were traditionally 
considered to be physically irrelevant 
(see however \cite{lavrentiev}).  This 
implies, for example, that there are problems with the physical 
interpretation of models  based on  GR with two timelike dimensions 
\cite{chaichian} (as also pointed out in  $(10+2)$--d 
supergravity \cite{Hewson}). 

If we exclude this second possibility, it may be more promising
to approach higher-dimensional embeddings with a
two--step procedure of the form 
$(3,0) \longrightarrow (4,0) \longrightarrow (4,1)$. The first
step could be a boundary problem 
along the lines considered in \cite{reula}, 
while the second step is a higher-dimensional version of the 
GR CP.
This procedure would allow the bulk to  
influence the hypersurface
in a causal way (in the usual GR sense),
which is not possible in a $(3,1) \longrightarrow (4,1)$ 
type of embedding  
(except perhaps in a perturbative treatment). It  
is clearly of interest 
to construct models in which the bulk does play a role that is, in principle,
observable and testable. Although the nature of causality in 
higher--dimensional theories may be substantially
different to that of $(3+1)$--d GR \cite{bwcausality}, 
such models would
be consistent with a direct extension of GR. 

\section{Conclusion}

In this paper, we have proved an extension of the CM embedding 
theorem \cite{campbell,magaard}, where the embedding metric is 
sourced by one or more 
scalar 
fields minimally coupled to 
GR. We employed this theorem to establish classes of embeddings 
where the scalar field self--interacts through an exponential potential. 
The relationship between the CM theorem 
and the Cauchy and initial value problems of GR was highlighted. 
This relationship will certainly permit CP an IVP techniques  
to be adapted to provide further space--time into space--time 
embeddings. 

Embedding theorems are important from both the mathematical and physical 
points of view. They allow classification schemes for different spacetimes 
to be developed as well as providing algorithms for generating 
new solutions \cite{mah}. 
The primary physical motivation for developing mathematical embedding 
theorems arises from the resurgence of interest in higher--dimensional 
theories of gravity and cosmology, most notably the braneworld scenarios \cite{BW,HW,lukas,RS}.
In particular,  
the Randall--Sundrum `type II' (RSII) scenario \cite{RS}, where a codimension 
one brane is embedded in five--dimensional Anti-de Sitter (AdS$_5$) 
space, has attracted considerable attention. 
Embeddings are also of relevance to the 
induced `space--time--matter' (STM) theory, whereby 
the 4-d universe is viewed as a slice of a 5-d universe
\cite{stm,PDL,searha}. (For reviews, see, e.g., Refs. \cite{ind1,ind2}). 
The CM theory was recently employed within the context of both 
the RSII scenario and STM theory 
where the close similarities between the two 
approachs were 
also highlighted and algorithms for employing the theorem in practice 
were outlined \cite{searha}. The relationship between the two pictures 
has also been emphasized in Ref. \cite{PDL}.

Although the RSII model was motivated in part by  
the Hor\v{a}va--Witten theory for the strongly 
coupled limit of the ${\rm E}_8 \times {\rm E}_8$ heterotic string 
theory \cite{HW}, it is an idealized system, in the sense that 
the higher--dimensional space is assumed to be an Einstein 
space. It is therefore important to 
consider more general settings that are inspired 
by string theoretic considerations. 
For example, the supergravity actions contain, 
in addition to the scalar dilaton field, a number of antisymmetric 
form--fields. Up on compactification from ten or eleven 
dimensions, the degrees of 
freedom associated with these fields can manifest themselves as scalar fields 
that self--interact through one or more potentials. In particular, 
scalar fields arise when the form fields have a non--trivial flux 
over the compactifying dimensions that is able to 
support a solitonic brane configuration. 
A specific example is the compactification of the Hor\v{a}va--Witten 
theory to five dimensions \cite{lukas}, where a scalar field arises in the fifth 
(bulk) dimension through an exponential 
potential of the form considered in Sec 3.2. 

In view of the above discussion, it is important to 
develop the mathematical framework for 
studying braneworld models containing scalar fields in the higher 
dimensions through embedding theorems. 
Unlike almost all\footnote{This is also 
the starting point of the perturbative 4--d treatment 
of domain walls presented in \cite{BSF}.}
current analyses, our treatment in the present work is for 
the 
{\em full}
Einstein--scalar field system rather than for some 
highly symmetric case. This general treatment is 
appropriate because we are establishing 
an existence and uniqueness theorem for 
embeddings into space--times sourced by scalar fields. 

It is worth emphasizing that 
the scalar fields we have considered in our extension of 
the CM theorem are additional, dynamical  
bulk scalar fields, 
$\bar{\chi}$, and not the higher-dimensional metric component, $\bar{\phi}$. 
The theorem we have proved therefore provides a starting 
point for considering a number of open physical questions, such as 
the 
possible restrictions that may arise on the energy--momentum tensor 
of the brane matter fields once the junction conditions 
are imposed. A related question is 
whether plausible 4--d physics can generically be recovered
and whether there are testable corrections to Newton gravity 
such as those in the RSII scenario \cite{RS}. 
It would be interesting to consider these questions further. 

Finally, our generalization in this paper of the CM theorem implies that
for a given 4-d space--time, there are an infinite 
number of embedding space--times with one extra dimension, in 
the sense that there is at least one possible embedding 
for each functional form of the higher-dimensional energy-momentum 
tensor\footnote{In the 
embedding procedure, the energy--momentum tensor is viewed as 
a function of the coordinates and not as a function of the fields.  
This implies that the higher-dimensional models are merely encoding solutions 
rather than specific physical laws.  Compare with how Kaluza--Klein
theory offered further predictions as a result of the geometrization 
of the electromagnetic field.}, 
in particular, one per scalar interaction 
potential of the surrounding 5-d space--time.  
This is arguably an undersirable feature for higher-dimensional theories, 
since
it implies that unique physical predictions can not be made in the absence 
of well-founded 
principles that {\it select} a particular higher--dimensional 
space--time.  
This is similar to the well--known problem in string theory 
phenomenology, 
where different particle spectra arise for different Calabi-Yau 
compactifications \cite{CY}.
The introduction of discontinuities or branes is unlikely to remove 
this non--uniqueness property, since there are
many possible brane configurations (single, parallel or intersecting) 
that could be considered  
and many possible forms for the higher--dimensional bulk space--time, 
but presently no established way to distinguish
between them from a physical point of view. Perhaps, 
the number of fields and the form of the potential would 
be restricted in some higher--dimensional theory and 
by requiring 
the recovery of (possibly corrected) 4--d physics as discussed above. 
Our theorem provides a framework within which these and 
related questions can be addressed. 

\vspace{.3in}

\noindent\bf{Acknowledgements}\normalfont

EA is supported by PPARC.  FD and CR are supported by CNPq.  JEL is supported by the Royal Society.  
We thank M. A. H MacCallum, N. \'{O}. Murchadha and R. Tavakol 
for helpful discussions and communications.

\end{document}